\documentclass[]{spie}  

\usepackage{amsmath,amsfonts,amssymb}
\usepackage{graphicx}
\usepackage[colorlinks=true, allcolors=blue]{hyperref}

\def\cfahs{{Center for Astrophysics \textbar\ Harvard \& Smithsonian}}

\def\3He{\textsuperscript{3}He}

\title{Laminate polyethylene window development for large aperture millimeter receivers}

\author[a]{Miranda Eiben}
\author[a]{Denis Barkats}
\author[a]{Aurelia Balkanski}
\author[a]{Sage Crystian}
\author[a]{Marion~I. Dierickx}
\author[a]{David C. Goldfinger}
\author[a]{Paul~K. Grimes}
\author[a]{Robert Kimberk}
\author[a]{John~M. Kovac}
\author[a]{Grant Meiners}
\author[a]{Matthew~A. Petroff}
\author[a]{Destiny Santalucia}
\author[a]{Elaine Sheffield}
\author[a]{Calvin Tsai}
\author[a]{Natalia Villanueva}

\affil[a]{\cfahs, 60 Garden St, Cambridge, MA 02138, USA}

\authorinfo{Send correspondence to Miranda Eiben, e-mail: miranda.eiben@cfa.harvard.edu}

\begin{document} 
\maketitle

\begin{abstract}
New experiments that target the B-mode polarization signals in the Cosmic Microwave Background require more sensitivity, more detectors, and thus larger-aperture millimeter-wavelength telescopes, than previous experiments. These larger apertures require ever larger vacuum windows to house cryogenic optics. Scaling up conventional vacuum windows, such as those made of High Density Polyethylene (HDPE), require a corresponding increase in the thickness of the window material to handle the extra force from the atmospheric pressure. Thicker windows cause more transmission loss at ambient temperatures, increasing optical loading and decreasing sensitivity. We have developed the use of woven High Modulus Polyethylene (HMPE), a material 100 times stronger than HDPE, to manufacture stronger, thinner windows using a pressurized hot lamination process.  We discuss the development of a specialty autoclave for generating thin laminate vacuum windows and the optical and mechanical characterization of full scale science grade windows, with the goal of developing a new window suitable for BICEP Array cryostats and for future CMB applications.
\end{abstract}


\section{INTRODUCTION}
\label{sec:intro}  
\subsection{Thin Windows}
Modern millimeter wavelength telescopes are designed around superconducting, and therefore cryogenic, detectors. Cryogenic temperatures require very low heat transfer to the detector stages, and thus a high vacuum to limit convective transfer. At the same time, the desired wavelengths of light must reach the detectors. This requires millimeter vacuum windows to be very transmissive to the desired wavelength bands but still strong enough to hold a vacuum.

BICEP Array is probing the millimeter bands at very high sensitivity to observe the degree scale B-mode anisotropies in the cosmic microwave background (CMB). The power distribution of the B-modes at the degree scale may be a direct signature of gravitational waves that resulted from inflation in the early universe. The BICEP Array receivers are modeled after the previous generation BICEP3 instrument, and many lessons learned from BICEP3 are carried into the design of BICEP Array. Every optical element is designed to minimize the systematic contribution to the CMB signal, which requires cryogenic lens to limit in-band emission \cite{Hui2018}. Instrument sensitivity can be increased by increasing sky coverage and detector count which, for a cryogenic refracting telescope like BICEP3 or BICEP Array requires a larger aperture size. Windows then become thicker to accommodate the increased atmospheric pressure load of a larger aperture, which results in larger transmission loss. 

A single BICEP Array receiver has an aperture of approximately 730mm (28.8in) in diameter, which results in a total atmospheric pressure on a window of 3000kg (6500lbs). The current window material used for BICEP3, High Density Polyethylene (HDPE), is 31.8mm thick to withstand the extreme atmospheric pressure and minimize center deflection. As the vacuum window is the only optical component that cannot be cooled, the current BICEP Array windows produce the largest contribution to the optical loading, with an estimated load between four and seven times that of other in-beam optical components \cite{B3paper}. The emissivity of the window in-band is directly related to the transmission attenuation (absorption) in band. Transmission attenuation loss goes with \(e^{-\alpha t}\), where t is thickness and \(\alpha\) is the attenuation coefficient of the material. In the low loss regime, reducing the window thickness linearly reduces transmission loss, and thus the emissivity of the window in band.

\begin{figure}
    \centering
    \includegraphics[width=0.7\textwidth]{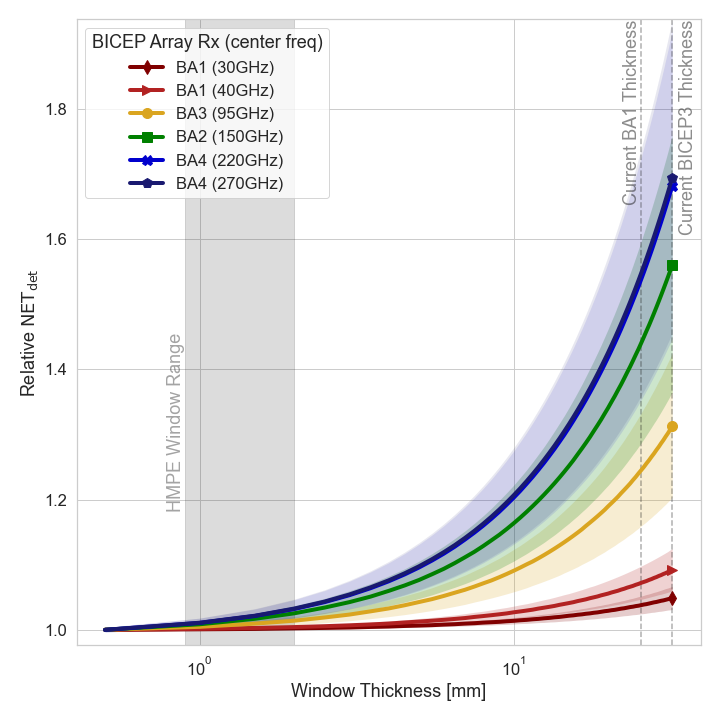}
    \caption{Per detector Noise Equivalent Temperature (NET) relative to a BICEP cryostat with no window, for each of the BICEP Array (BA) bands. Colored bands represent transmission loss uncertainty of polyethylene between 2e-4 and 4e-4 \cite{Lamb1996}. Lines indicate current polyethylene window thicknesses on BICEP cryostats, which are also representative of the bands planned for CMB-S4. Shaded region indicates the range of goal thicknesses for thin windows.}
    \label{fig:NET}
\end{figure}

Decreasing the window thickness (while continuing to use low loss materials) would decrease the thermal noise on the detectors, resulting in faster mapping speeds. Noise on the detectors of CMB instruments are typically described with the noise equivalent CMB temperature on a detector (NET$_{\text{det}}$), which takes into account the various noise sources in a millimeter receiver such as the photon noise incident on a detector, phonon noise in a detector, and readout noise from a detector. Figure \ref{fig:NET} summarizes the theoretical reduction in NET$_{\text{det}}$ for each of the BICEP Array bands when the vacuum window thickness is reduced from the current BICEP3 (31.8mm) and 30/40 GHz BICEP Array receiver (BA1) (25.4mm) thicknesses. A solution to the opposing goals of vacuum windows can be found by using a stronger material for the vacuum window, which allows a decrease in thickness without compromising the mechanical integrity of the vacuum window. Using High Modulus Polyethylene (HMPE), which is a factor of 100 stronger than conventional High Density Polyethylene (HDPE), for windows is a potential solution: strong enough to hold a vacuum while remaining very thin to reduce noise on our detectors. 

Polyethylene (PE) is an common thermoplastic polymer commercially available in a variety of forms. It is chemically resistant and stable, and very transmissive in the microwave \cite{Vasile2005,Lamb1996}. The high transmission and low index of refraction, plus wide commercial availability, leads to PE being used in many microwave instruments for optics such as windows or lenses \cite{B3paper,SeanBryan2018,Thornton2016,Swetz2010,QUIETCollaboration2012,Rhoades2018,Essinger-Hileman}. By utilizing different fabrication processes, the long complex molecular chains of the polymer can have different shapes. For example, two common forms of PE include the branched entangled structures of Low Density Polyethylene (LDPE) and the denser single entangled chains of High Density Polyethylene (HDPE) result in differing material properties, such as density, ductility and tensile strength. Entangled polymer chains allow for extensive deformation under stress, but very little force is required for the chains to unwind \cite{Vasile2005}.

High Modulus Polyethylene (HMPE) is a form of polyethylene wherein the the long complex molecular chains are aligned into fibers in a gel spinning process instead of being allowed to tangle. This allows the material to have a tensile strength to weight ratio comparable to steel.  It is sold as a rope or woven fabric \cite{Dyneema}.  A woven fabric, however, is not vacuum tight.

Experimentation has shown that high pressure lamination is the key to make a vacuum tight HMPE window. We have developed a lamination process with LDPE. LDPE, due in part to shorter polymer strands, has a lower melting point ($T_{m,\text{LDPE}}$= 104-114$^{\circ}$C, $T_{m,\text{HMPE}}$ 145-155$^{\circ}$C) \cite{MahdaviHossein2008Cams,Dyneema} and flows well when melted \cite{Vasile2005}. This makes it an ideal laminating or matrix material to flow between the fibers of HMPE and to generate a composite material that appears homogeneous to microwave wavelengths.

Small scale testing, however, has indicated that laminates generated at low pressures leak excessively. We suspect this is due to improper lamination around the grain (or fibers) of the HMPE. Gas can follow the fibers to then penetrate the vacuum either by permeating through laminating layers or through micro-voids in the plastic. High pressure lamination is the key to force the laminating material around the grain and prevent gas from flowing along it.

\subsection{Anti-Reflection Coating for Large Apertures}

\begin{figure}
    \centering
    \includegraphics[width=0.8\textwidth]{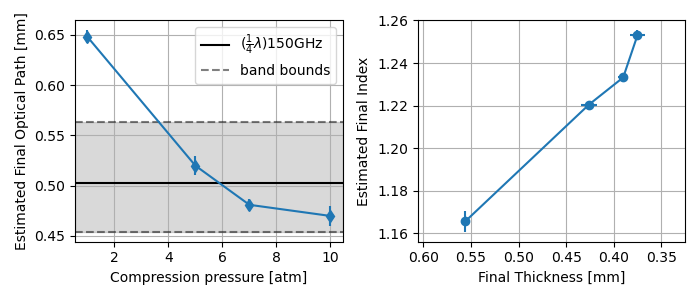}
    \caption{Optical path lengths vs compression pressure (left) and final indexes and thicknesses (right) of compressed ePTFE. }
    \label{fig:ar_comp}
\end{figure}

Anti-reflection (AR) coatings are vital to reduce the power lost from reflections and spurious signal from being reflected onto detectors. By reducing the impedance mismatch between air or vacuum with a material that is designed to cause destructive interference to reflected light, reflected power off a PE optic can be reduced from 20\% at certain frequencies in band to less than a percent. As aperture size grows it is increasingly difficult to source continuous AR coatings for polyethylene optics at millimeter wavelengths at the scales required. Current small aperture telescopes (SAT) observing the CMB use either polyethylene lenses (most of the BICEP/Keck series \cite{BICEP1_2010,BICEP2_2010,Keck2010,Schillaci_2020}), alumina (BICEP3 \cite{B3paper}), or silicon (CLASS \cite{CLASS2018}). Each lens material has advantages and disadvantages; one of the advantages of PE lenses is that the low index of refraction allows for easier index matching for ideal anti-reflection coatings. Polyethylene lenses were chosen over the other options for future BICEP Array receivers due to the availability of cast PE at large scales (not available for materials like lab grown silicon), ease of machining (unlike brittle materials like alumina), and an optics redesign that allowed for thinner PE lenses (and therefore lower absorption) than previous PE designs.

 With lenses and windows getting increasingly large, sourcing previous AR coatings for PE such as sintered PTFE (sPTFE) in continuous sheets at the correct densities and thicknesses is difficult and expensive. We have turned to expanded PTFE (ePTFE) which utilizes an expansion technique to stretch the PTFE into a lower density membrane; however, this expansion introduces thickness and density variations that result in optical path variations of up to 15\% across a single sheet.

We have developed a heat-compression technique to control the thickness and index of expanded PTFE post-factory production which also reduces property variability in large ePTFE sheets \cite{Dierickx2021}. The compression pressure can be varied to control the final optical path length of the AR coating, while the temperature is increased above a second order phase transition of the PTFE known as the glass transition temperature and then cooled under pressure, to allow the compression to be permanently locked in \cite{Calleja2013}. Figure \ref{fig:ar_comp} shows the relationship that compression pressure has to the optical path length of an ePTFE sheet to be used for AR coating 150 GHz PE optics. 

In the rest of these proceedings we will discuss the following: the pressure oven we developed to heat compress both laminate windows and ePTFE AR coatings (Section \ref{sec:autoclave}), including the mechanical and thermal characterization of the machine; the optical, vacuum, and mechanical results of thin laminate windows produced in the previously described autoclave (Section \ref{sec:results}); finally the future work required to finish characterization of thin laminate windows (Section \ref{sec:fut_work}).

\section{OPTICS AUTOCLAVE}
\label{sec:autoclave}
The lamination and AR coating compression processes described in Section \ref{sec:intro} require the use of a heated press large enough to contain the complete optical component to be processed.  The lamination processes require temperatures up to 140$^{\circ}$C and pressures up to 10 atm. In order to provide uniform pressure over optical elements, we place the optical elements in a vacuum bag, and apply pressurized gas around the bag to provide the necessary compression for lamination.

Thin laminate windows require pressures at least above two atmospheres, more safely above four, to fully laminate the HMPE fibers. They require a specific temperature range between the melting point of LDPE (104-114$^{\circ}$C)\cite{MahdaviHossein2008Cams} and the melting point of the HMPE (145-155$^{\circ}$C)\cite{Dyneema}. Small scale tests have shown that adequate lamination occurs between 130 and 140$^{\circ}$C, with the best range between 133 and 137$^{\circ}$C. Deviations outside those temperatures can result in inadequate or uneven lamination of the fibers, creating an unusable window.

The ePTFE heat compression requires a range of pressures, depending on the material the AR coatings are designed for. The temperature needs to be adequately above the glass transition temperature of PTFE, which is at maximum estimated to be 120$^{\circ}$C\cite{Calleja2013}.

These two use cases put similar requirements on the autoclave, as summarized in Table \ref{tab:requirements}.

\begin{table}
    \caption{Temperature and pressure requirements for thin laminate windows and expanded PTFE anti-reflection coatings  for both polyethylene optics (windows and lenses) and nylon (infrared filters).}
    \centering
    \begin{tabular}{|c|c|c|}
        \hline
\rule[-1ex]{0pt}{3.5ex} \textbf{Material} & \textbf{Temperature} & \textbf{Pressure}  \\
        \hline
\rule[-1ex]{0pt}{3.5ex} Thin Windows & 130-140$^{\circ}$C & $>$4atm (207 kPa)\\
        \hline
\rule[-1ex]{0pt}{3.5ex} ePTFE (PE and Nylon AR coating) & $>$130$^{\circ}$C & 1-10atm (101 – 1,013 kPa) \\
\hline
    \end{tabular}
    \label{tab:requirements}
\end{table}

\subsection{Design}
\begin{figure}
    \centering
    \includegraphics[width=\textwidth]{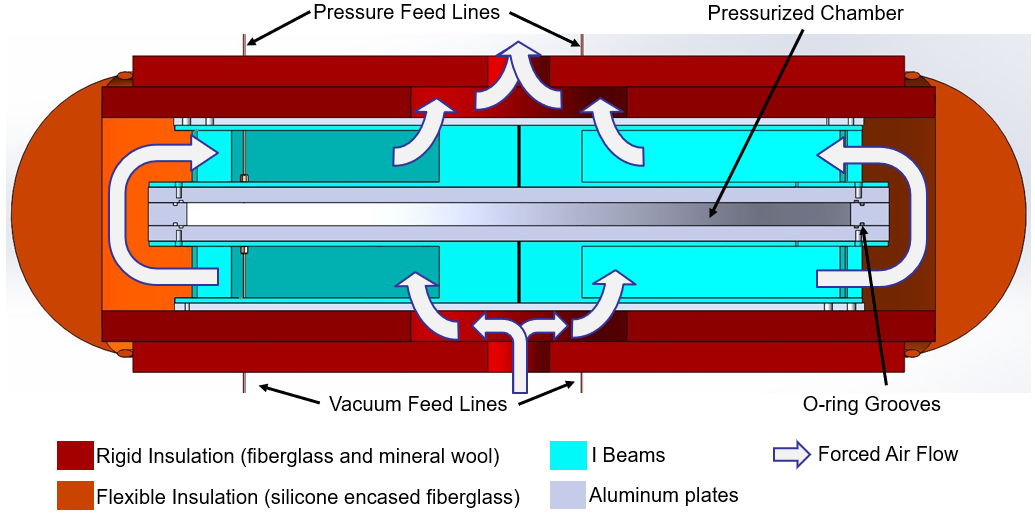}
    \caption{Cut away diagram of autoclave. Warm colors (red, orange) indicate insulation, cyan are I-beams, grey are machined aluminum plates. Arrows with blue borders indicate forced air directions.}
    \label{fig:my_label}
\end{figure}

\subsubsection{Mechanical}

The scale of the optics materials that need to be compressed is currently at most 914mm (36") in diameter and at most a few millimeters thick. This allows for a very narrow pressurized chamber for compression. The combination of the large size and pressure required for compression results in a very large force that must be contained with the top and bottom plates subject to approximately 1 million Newtons (266,000 lbs or 120,655 kg) of force at the maximum required pressure. 

One inch thick aluminum plates would not be able to safely withstand this force, but thicker plates would quickly become prohibitively heavy; the maximum load on our hoists is one ton (800 kg) and the current autoclave weighs 400 kg. To reduce weight while adding strength, we supplemented the one inch aluminum plates with aluminum I-beams arranged in a radial pattern and around the circumference of the plates. The I-beams and plates are bolted together with 1/2" steel bolts, with cast iron and cast aluminum washers to mate the bolts and the angled I-beam surfaces. The load is further redistributed among the I-beams with a `wagon wheel' aluminum plate above the I-beams. 

To prevent bubbles, wrinkles or movement of layers during the compression process, the materials are placed inside a vacuum bag and pumped on continuously during the oven cycle. Additionally, the entire interior chamber is pumped down with the vacuum bag to aid in gas escaping underneath the bag. Then the chamber above the bag is pressurized to the desired pressure.

The plates are mechanically coupled with a two and a half inch wide and one and a half inch thick outer ring, and 48 Grade 8 3/4" bolts. The bolts run through both plates and the outer ring, and are secured with Grade 8 nuts underneath the bottom plate. The outer ring allows for two silicone o-rings to be used to seal the chamber between the two plates. We have found that the weight and force of the bolts is sufficient to hold pressure with only two o-rings above and below the outer ring, though there is the option of additional o-rings to isolate the vacuum bag further from the pressurized chamber. The optical materials are placed under a flexible vacuum bag material and the outer ring and top plate are placed over the vacuum bag.  

\begin{figure}
    \centering
    \includegraphics[width=0.8\textwidth]{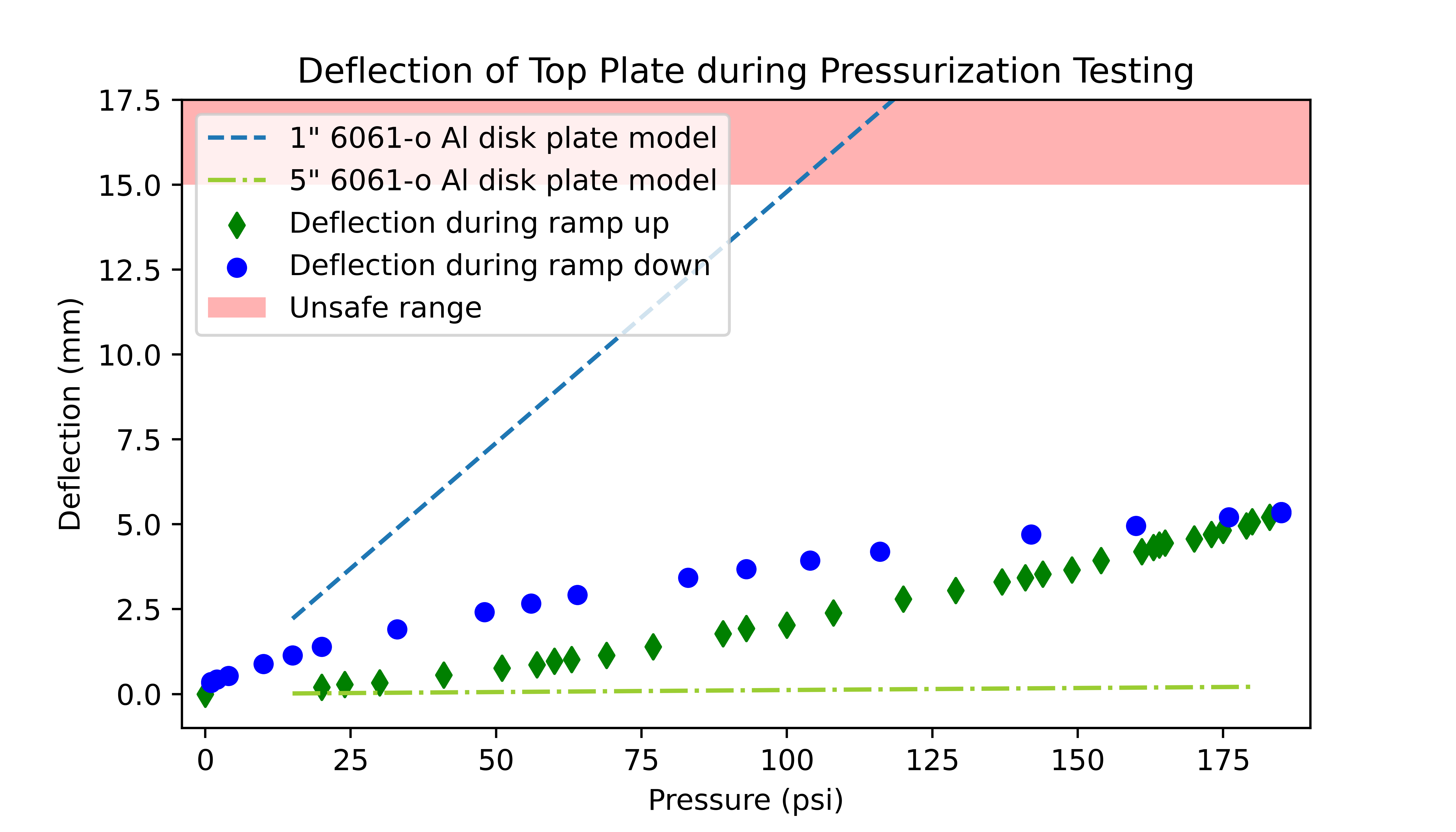}
    \caption{Center deflection measurements during initial hydrostatic pressure testing. Deflection showed some hysteresis between the ramp up (diamonds) and ramp down (circles) but remains within safety margin. Dashed lines show expected aluminum plate deflection for a 1" plate (dash-dotted green line) and a 5" plate (dashed blue line).}
    \label{fig:press_def}
\end{figure}

We tested the deflection of the top plate when the chamber was hydrostatically loaded up to 1.276 MPa (25\% above the maximum required load). Results are reported in Figure \ref{fig:press_def}. There was some hysteresis in the deflection between ramp up and ramp down, but both stayed between the calculated deflections of a 1" and 5" thick plate, and well below the upper limit requirement of 15 mm. 

Pressure is achieved during heat compression with pressurized nitrogen gas. The pressure is regulated at the canister with a regulator, and monitored at both the canister and the chamber with analog dials. A back pressure regulator acts as a pressure release valve to prevent over pressure at temperature. 

\subsubsection{Thermal}
Thermal insulation on the autoclave is a mix of rigid top and bottom insulation, with a flexible outer wrapping. The rigid insulation used is both mineral wool and fiberglass sheets, which have similar thermal resistance. The flexible insulation is a silicone encased fiberglass fabric, sewn into shape with Kevlar thread, around a core of fiberglass wall insulation. 

A total of 48 heater strips are used, each capable of producing 240 Watts of heat per heater. The oven requires a total of 9 kW to reach the set temperature. These heaters are attached to the vertical face of each I-beam, with one heater occupying each I-beam. The heaters are mechanically and thermally connected to the I-beams with two screws on either end of the heater. Further thermal conduction is aided by the addition of thermal compound Chemax Tracit-1100 between each heater and the I-beam. The heaters along the top and bottom halves of the autoclave are wired in parallel to the other heaters on the same half. Thermal fuses rated to 170$^{\circ}$C are also included in series with the heaters to trip in the event of overheating. The heaters are controlled via a PID loop with three thermometers per side at two different radii to maintain temperatures as specified by the recipe for the particular optics element that is being produced. The integral term of the PID has an upper limit to prevent windup, and the hardware has an additional watchdog timer to cut the output after a few seconds if the control script fails to send an update.

Forced air flowing from below the autoclave, through the I-beams and out the top insulation acts as a thermal regulator. When the autoclave is heating or at temperature the air is recycled in a closed loop, and when cooling the loop is broken by removing the duct connecting the top and bottom. The heating cycle takes approximately three hours, then the autoclave is held at temperature for two hours to ensure the interior temperature is isothermal and reaches the set temperature. Cooling from the set temperature takes approximately eight hours.

\begin{figure}
    \centering
    \includegraphics[width=0.45\textwidth]{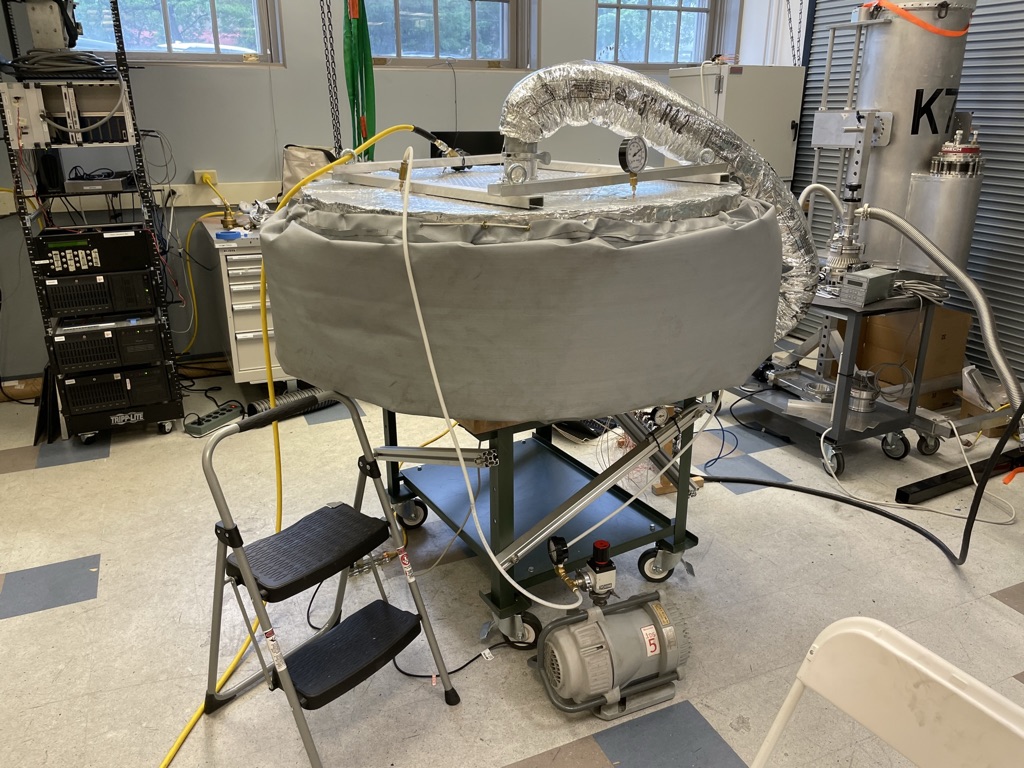}
    \includegraphics[width=0.45\textwidth]{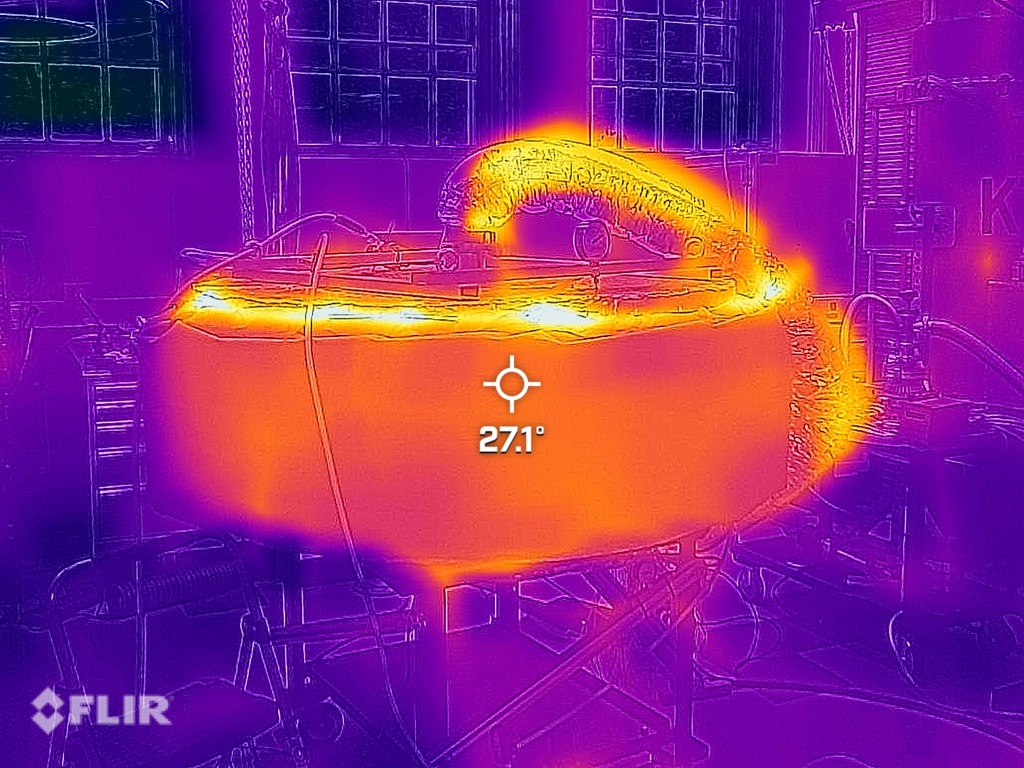}
    \caption{Optical and thermal image of the autoclave in operation. Interior temperature is set to 135$^{\circ}$C.}
    \label{fig:op_therm}
\end{figure}

\section{RESULTS}
\label{sec:results}
\subsection{Optical}

The optical performance of an optic is dependent on the reflections off the interfaces and the loss within the material. Reflections can be minimized with appropriate anti-reflection coating, and loss can be minimized with material choices and the thickness of the material. Tuning the thickness of the thin window along with the compressed thickness and refractive of the ePTFE AR coatings allows multiple reflection nulls to be generated within the bandwidth of the receiver, allowing for low reflections over a relatively wide bandwidth.

We measured reflections off of sample AR coated windows on a on a millimeter-wave Vector Network Analyzer equipped with W-band (95 GHz) and D-band (150 GHz) frequency extenders in a single port configuration. Each waveguide frequency extension VNA head is used with a rectangular 23 dB standard gain horn and 90° offset parabolic mirror to produce a Gaussian beam waist at the sample position.  Samples are placed at the Gaussian beam waist perpendicular to beam direction, and the reflection from the sample compared to that from a machined aluminum plate, which is assumed to have near perfect reflectivity. The beam transmitted through the sample is terminated some distance behind the sample on an angled Eccosorb HR-10 absorber, and time gating of the VNA reflection signal is used to eliminate spurious reflection signals from the horn, collimating mirror and beam termination.

 We measured the reflection out of band in the W-band (75 – 110 GHz) to get a higher signal measurement for a window meant for the 150 GHz (133 – 160 GHz) BICEP Array receiver. Two samples were made to compare the previous small autoclave used for prototyping and the final autoclave. Both AR coating and laminates were compressed in the separate autoclaves. The samples produced essentially identical reflections in the W band, and minor variations may be explained by positioning uncertainties.
 \begin{figure}
    \centering
    \includegraphics[width=\textwidth]{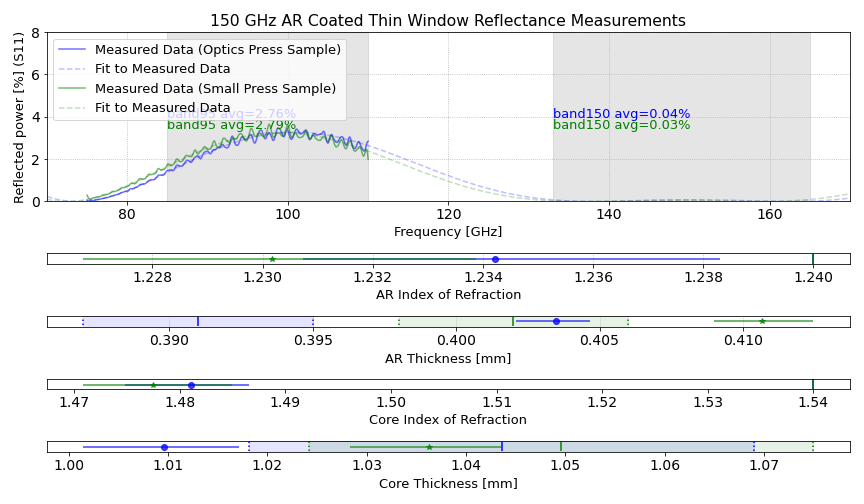}
    \caption{[Top] W-band vector network analyzer (VNA) S11 measurements of thin window samples anti-reflection coated for the 150 GHz band. [Bottom] Fit parameters and uncertainties to the reflection measurements. Solid lines are starting parameters; dotted lines are measurement uncertainties on thicknesses.     }
    \label{fig:vna_results}
\end{figure}

We used the generative model with the transfer matrix method to fit to the VNA data with a uniform prior between a reasonable maximum and minimum index on all index parameters and a normal prior with measurement uncertainties on all thickness parameters. The unnormalized likelihood function is assumed Gaussian, as we expect the noise on the reflected return to be Gaussian due to dominant positioning effects. The uncertainties on the data were determined from the maximum differences between reference planes (or shorts), to constrain the potential minor positional or produced power uncertainties and differences between measurements. Samples were rotated 90$^\circ$ between measurements, to constrain orientation and positioning effects on reflected power.

\subsection{Vacuum}
\label{sec:vacuum_res}
Permeation of gas through materials is inversely proportional to the thickness of the material. By significantly decreasing the thickness of the vacuum window by 95\%, we could be significantly increasing the amount of gas entering our cryogenic vacuum chamber, which could adversely impact the cryogenic performance of the cryostat. To test the potential impacts of this permeation of gas, we both leak checked the first prototype thin window and measured the gas entry rates into the cryostat.  

The first thin window prototype from the optics autoclave was helium leak tested on a small vacuum chamber for a few hours. There was no noticeable jumps in leaks into the chamber during that time, though there was a slow background increase over a few minutes. A slow helium increase in the vacuum is associated with permeation through the membrane; therefore, we decided that the window was leak tight. The window was then added to a local BICEP Array cryostat for a month-long calibration run.

Pressure was monitored on the BICEP Array cryostat continually for the month that the cryostat was cold. While cold the cryostat had no noticeable increases above 1e-5 mbar, which is the lower limit of the pressure transducer (an MKS 925). This is expected, given that the cryostat was cold and thus cryo-pumping on any gas that entered the chamber. There were also no thermal anomalies that previous BICEP cryostats have seen that are associated with gas undergoing phase transition and absorbing heat off cold components. However, this calibration run was significantly shorter than the typical observation season, so it is unlikely that sufficient amounts of gas could accumulate to cause thermal anomalies.

If we take the final pressure of the last two runs of the BICEP Array cryostat, we can compare the measured average gas entry rates into the vacuum. By assuming that the temperature of the gas is consistent, we can estimate the amount of gas in the cryostat with the final pressure in the cryostat and the volume of the interior of the cryostat. If we assume that all of the gas entered into the cryostat at a steady rate during the time at vacuum, we can estimate a measured average gas entry rate into the cryostat.  

\begin{table}
    \caption{Gas entry rates measured in a BICEP Array cryostat during short calibration runs in Cambridge, MA (at sea level). BICEP Array cryostat inner volume is 905 L. }
    \centering
    \begin{tabular}{@{}|c|c|c|c|@{}} 
    
        \hline
          & \textbf{Time at Vacuum} & \textbf{Final Warm Pressure} & \textbf{Meas. Av. Gas Entry} \\ 
         & [days] & [mbar] & [1e-6 $\frac{\text{L mbar}}{\text{s}}$]\\ 
         \hline
         1.2mm Window & 30 & 0.85 & 290 \\
         \hline
         25.4mm Window & 80 & 1.2 & 175 \\ 
        \hline 
        
    \end{tabular}
    \label{tab:ge_rates}
\end{table}

The direct comparison of these two runs shown in Table \ref{tab:ge_rates}, ignoring the potential confounding differences between them, suggests that the inclusion of the thin window increased the gas entry rate of the cryostat by an additional 65\%. However, these tests were performed at sea level and the operating location is at South Pole (9,300ft, 0.71 atm) and therefore we should expect that the permeation rate at South Pole would be decreased by approximately 30\%. However, there are many potential confounding factors to this rough estimate; for example, the amount of time that the cryostat stays on a pump before the cool down starts could reduce the amount of gas adsorbed onto inner surfaces before they are frozen to them. Additional components added to the cryostat between runs may outgas slowly and cause an initial increase in the warm pressure after a short run and disappear in subsequent runs. Despite this, the final pressures between these runs and their measured gas entry rates are remarkably similar. This lends credence to the interpretation that the addition of a thin window did not and will not adversely impact the cryogenic performance of the instrument. 

\subsection{Mechanical}

Plastic materials respond to continuous stress differently than other solids. Instead of an initial deflection under load that remains static, plastics will creep logarithmically over time. The initial elastic deformation of the window can be predicted reasonably well by flat plate predictions from Theory of Plates and Shells \cite{TimoshenkoStephen1959Topa} as

\begin{align}
	\delta = K \Big(\frac{\Delta P \, R^4}{E \,t}\Big)^{(1/3)} \label{eq:delta} 
\end{align}
where K is a constant depending on the location of interest of the plate (K = 0.662 for the center), $\Delta$P is the pressure difference (assumed to be atmospheric), R is the radius of the plate (39cm) and t is the thickness of the plate (1.2mm). Assuming that the minimum elastic modulus (E) of the window is 1.39 GPa (found from previous tensile strength tests), we can make a prediction of the initial deflection of the window without creep.

The strain of the window is estimated assuming the deflection of the window is radially symmetric, and that the deflected window is parabolic. We can then calculate the arc length of the parabolic window from the deflection of the window. Therefore, the strain of the window is calculated as

\begin{align}
    \text{Radial strain} = \varepsilon = (R_{\text{arc}}-R_{i})/R_{i}\label{eq:rstrain1}\\
    R_{\text{arc}} = \frac{\sqrt{R_{i}^2+4\delta^2}+(\frac{R_{i}^2}{2\delta}\sinh^{-1}[{\frac{2\delta}{R_{i}}}])}{2}\label{eq:rstrain2}
\end{align}
where $R_{i}$ is the initial radius and $\delta$ is the center deflection. We fit to this strain with a power law relation similar to those described in Ref.~\citenum{FindleyWilliamNichols19741cop} for polyethylene over a sixteen year long strain measurement.

\begin{figure}
    \centering
    \includegraphics[width=0.8\textwidth]{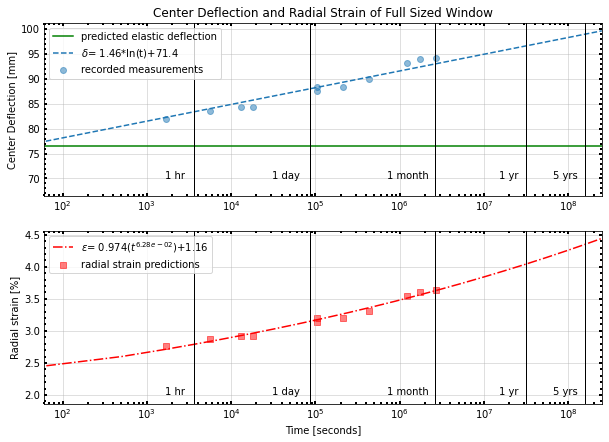}
    \caption{[Top] Measured deflections of a full sized thin window on a cryostat (blue circles), with a logarithmic fit to the deflection (dashed blue line). [Bottom] Estimated radial strain of the window using deflections and Equation (\ref{eq:delta}) (red squares), with a power log fit to the strain (dash dotted red line).}
    \label{fig:creep}
\end{figure}

We performed verification of the creep rate of a prototype thin window without ePTFE AR-coating on a BICEP Array cryostat. We used the same window clamping frame that we have used previously for deployed BICEP Array HDPE window. The frame clamps the window between two rings with knurled surfaces that dig into the window to grip it; therefore, we assume that the outer is a clamped edge. As the cryostat was pumping out and cooling, we performed multiple measurements of the deflection with a flat metal meter stick and measuring the deflection with a right angle ruler by hand. During the initial helium leak tests discussed in Sec. \ref{sec:vacuum_res}, the thin window initially deflected 73 mm which is close to the predicted deflection from Equation (\ref{eq:delta}) of 76.5 mm. After one month under vacuum and the cryostat being continuously cold, the final deflection measurement prior to venting the cryostat was 94.28mm. 

The predicted strain of the window is staying below the previously found strains at failure during strength testing of between 10 to 15\%, even when we extrapolate out to years long time scales. However, we do not have the full scale, long term measurements to confirm that these models will hold on on the operational timescales of years. 

\section{FUTURE WORK}
\label{sec:fut_work}
\begin{figure}
    \centering
    \includegraphics[width=0.45\textwidth]{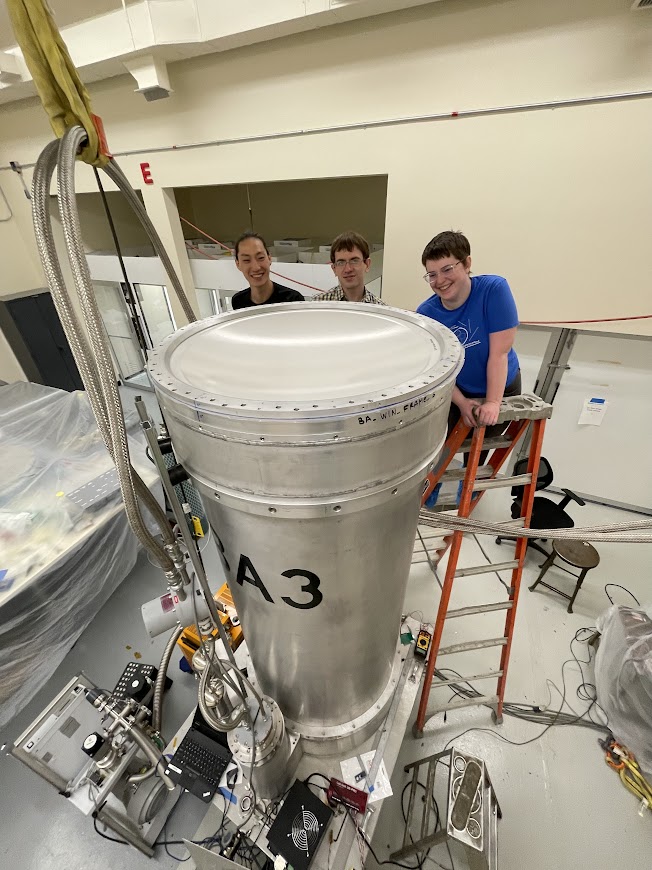}
    \caption{The first high pressure laminated thin window on a BICEP Array cryostat. People for scale.}
    \label{fig:window_cryostat}
\end{figure}
Currently, preparations for deploying two thin laminate windows to South Pole are underway. One is planned to replace the current slab window on BICEP3 and the other as the primary science window for the 150 GHz BICEP Array (BA2). A prototype of the BA2 science grade window is on a BA style receiver is shown in Figure \ref{fig:window_cryostat}. Further work will include optimization of such windows for broader frequency performance across the frequency bands planned for future BICEP Array and CMB-S4 receivers and potentially other millimeter-wave instruments.  These efforts and tests conducted previously on the performance of laminate windows including permeation of gas, thermo-mechanical effects under vacuum, scattered power, polarization upper limits, and many other characterization measurements will be reported in more detail in a future publication.

At this point, the laminate windows have been well characterized mechanically, with both strength tests and short term creep tests reported both here and in Ref. \citenum{Barkats2018}. The creep response should be tested on longer time scales to ensure the creep rate model and strain models remain consistent on the years long deployment time scales. Additional full scale stress tests may include destructive drop tests; we have previously confirmed that small tools likely to be dropped such as hex keys do not cause failure but have not tested dropping items to the stress limit of a window.

Optically, there are a few unknowns that remain. The transmission loss of the window has not yet been measured precisely, and we are currently in the process of developing a quasi-optical open resonator system to enable precise loss measurements of materials and optical components. The scattered power from a window has been constrained but not directly measured. Radiometric testing of the combined scattering and absorption loss of a full scale window may be carried out by placing a window in front of an operating receiver during a sky dip. Further constraints on the birefringence of the laminates may also be attempted. Measured changes of NET may be possible this coming austral summer if the planned replacement of the BICEP3 window occurs. 

\section{CONCLUSIONS}
We have demonstrated that large aperture thin vacuum windows can be produced by laminating woven HMPE with continuous LDPE sheets by heating above the LDPE melting point while under pressure. Additionally, large scale ePTFE AR coatings for polyethylene and nylon require heat compression to reduce optical path variability across the sheet and optimize the final thickness and refractive index. These use cases have led us to develop a large optics autoclave.

The optics autoclave has been proven to produce thin windows and AR coatings that preform well optically. The autoclave was tested hyrdostatically up to 125\% of the required pressure of 10 atmospheres, and controls the temperature of the interior to within one degree Celsius. Anti-reflection coatings and windows produced in the autoclave have been proven to perform as expected optically, mechanically and under vacuum. We plan to deploy thin laminate windows on the next telescopes sent to South Pole for the BICEP Array, and replace the BICEP3 slab window in the coming austral summer.

\appendix    

\acknowledgments 
 
Thanks go to Dr. David Krider for his experimental contributions in laser cutting these laminates. Thanks to Harvard University and \cfahs  for providing funding support of this work.

Many thanks to the generous work provided by undergraduate students and instructors, including those participating in Astronomy 191 at Harvard. Their labor has proven invaluable for the completion and characterization of these materials.
\bibliography{report} 
\bibliographystyle{spiebib} 

\end{document}